\documentclass[%
 reprint,
 longbibliography,
 amsmath,amssymb,
 aps,
]{revtex4-2}
\usepackage{CJK}
\usepackage{graphicx}% Include figure files
\usepackage{dcolumn}% Align table columns on decimal point
\usepackage{bm}% bold math
\usepackage{gensymb}
\usepackage{braket}  
\usepackage{booktabs}
\usepackage{makecell}
\usepackage[colorlinks=true, linkcolor=blue, citecolor=blue, urlcolor=blue]{hyperref}% add hypertext capabilities

\begin{document}
\begin{CJK*}{UTF8}{gbsn}

% \preprint{APS/123-QED}

\title{Compact Continuous Cold Atomic Beam from a Single Cell with 3D Cooling and Ultra-low Light Shift}

\author{Sheng-Zhe Wang (王圣哲) \(^{1,2}\)}
\author{Qian-Lan Cai (蔡千澜) \(^{1,2}\)}
\author{Zhi-Xin Meng (孟至欣) \(^{3}\)}
\author{Yi-Cheng Deng (邓意成) \(^{3}\)}
\author{Yan-Ying Feng (冯焱颖) \(^{1,2,}\)} 
\email{yyfeng@tsinghua.edu.cn}

\affiliation{
\(^{1}\) A-Knows Lab, Department of Precision Instruments, Tsinghua University, Beijing 100084, China}

\affiliation{
\(^{2}\) State Key Laboratory of Precision Measurement Technology and Instruments, Tsinghua University, Beijing 100084, China}

\affiliation{
\(^{3}\) Beijing Institute of Aerospace Control Devices, Beijing 100854, China}

\date{\today}% It is always \today, today,
             
\begin{abstract}

We report a compact single-cell source of a continuous cold-atom beam with three-dimensional (3D) cooling. By integrating an off-axis moving optical molasses (OM) with a two-dimensional magneto-optical trap (MOT), we achieve simultaneous 3D cooling within a $50~\mathrm{mm}$ interaction region. The source delivers a continuous flux up to $4.9(5)\times10^9~\mathrm{atoms/s}$, with a transverse temperature of $94(5)~\mu\mathrm{K}$, a longitudinal temperature as low as $231(65)~\mu\mathrm{K}$, and a tunable mean velocity between $5$ and $20~\mathrm{m/s}$. Custom in-vacuum mirrors integrate the reflective geometry for the off-axis OM beams with a $0.8~\mathrm{mm}$ output aperture, ensuring stable alignment while suppressing stray light and fluorescence leakage. Ultra-low light shift and decoherence are verified via continuous Raman-Ramsey interferometry, yielding a light shift of $-0.51(4)~\mathrm{Hz}$ and a typical fringe contrast of $90.85(30)\%$ at a Raman separation of $100~\mathrm{mm}$ (interrogation time of $8.70~\mathrm{ms}$). This compact continuous cold-atom beam source constitutes a practical building block for atomic-beam clocks and interferometers, enabling reduced aliasing noise together with improved sensitivity and accuracy for field applications.

\end{abstract}

\maketitle
\end{CJK*}

\section{\label{sec:level1} Introduction}

Continuous cold-atom beam sources have attracted broad interest across diverse applications, including atomic frequency standards \cite{Devenoges2012,Feng2015,Elgin2019,Biedermann2013}, inertial sensing with atom interferometry \cite{Xue2015,Kwolek2022,Meng2024}, quantum computation and simulation \cite{Schmiedmayer2002,Bluvstein2024,Chiu2025}, atomic lithography \cite{Behringer1996}, Bose-Einstein condensation \cite{Chen2022}, and precision spectroscopy \cite{Domenico2001}. Such beams combine low mean velocity, narrow velocity distribution, and continuous operation. For clocks and interferometers, low velocity extends interrogation time within a compact geometry, while a narrow velocity distribution enhances fringe contrast and sensitivity. Continuous operation eliminates the dead time inherent to pulsed sources, thereby suppressing aliasing noise from undersampling \cite{Dick}.

Two main strategies have been pursued to realize continuous cold-atom beams. Early demonstrations extracted atoms from effusive thermal beams \cite{Phillips1982,Ertmer1985,Zhu1991,Ketterle1992}. In particular, two-dimensional (2D) optical molasses (OM) were used to transversely cool effusive beams for interferometry \cite{Gustavson1997,Yan2023,Sato2025}. These sources deliver high flux but retain large mean velocities and lack longitudinal cooling, limiting their utility for compact, high-contrast interferometers.

An alternative approach generates cold beams directly from vapor cells using magneto-optical traps (MOTs), with atoms continuously extracted by unbalanced radiation pressure. Depending on field configurations, typical schemes include the low-velocity intense sources (LVIS) or three-dimensional (3D) MOTs \cite{Lu1996,Wang2003,Wang2011}, 2D MOTs \cite{Schoser2002,Kellogg2012}, and 2D$^+$ MOTs \cite{Dieckmann1998,Arlt1998,Camposeo2001,Kohel2003,Xie2022,Park2012,Wang2024}. These architectures achieve low transverse and longitudinal temperatures but often suffer from reduced flux compared to thermal beam sources. Variants employing 3D OM or 2D MOTs assisted by moving OM have been demonstrated \cite{Weyers1997,Berthoud1998,Berthoud1999}, though with even more modest flux.

In most MOT-based sources, pushing laser beams and near-resonant fluorescence leaking along the extraction axis induce light shifts (also known as ac Stark shifts) and decoherence in the downstream interrogation cell \cite{Dutta2016,Savoie2018}, thereby limiting precision applications. Hollow cooling laser and thin pushing laser beams in 2D MOT geometries have been used to reduce the near-resonant light, but this design still results in a residual light shift of about $-200~{\rm Hz}$ \cite{Huang2016}. Other methods spatially separate the atomic trajectory from the leaked light using a light trap, gravity, or multi-stage cooling \cite{Domenico2011,Xue2015,Kwolek2020}. However, gravity is ineffective for instruments operating under dynamic conditions or in free fall, while the light trap and multi-stage cooling increase the system complexity.

Single-cell MOT-based sources typically provide sub-Doppler transverse cooling, but their longitudinal temperature remains on the order of tens of millikelvin \cite{Dieckmann1998,Arlt1998,Camposeo2001,Schoser2002,Kohel2003,Xie2022,Park2012,Wang2024,Kellogg2012,Wang2003,Wang2011}. A broad longitudinal velocity distribution reduces interferometric contrast, especially under high rotation rates where velocity-dependent phases or scale-factor instabilities accumulate \cite{Black2020,Narducci2022}. Continuous multi-stage cooling approaches 
provide further longitudinal cooling \cite{Domenico2011,Kwolek2020}, but compactness remains a central challenge for deployable systems.

In this work, we demonstrate and characterize a compact single-cell continuous atomic-beam source with simultaneous 3D cooling. By combining a 2D MOT with an off-axis moving OM (two pairs of counterpropagating laser beams intersecting at an angle relative to the extraction axis), we obtain a high-flux beam with significantly reduced longitudinal temperature compared to conventional MOT-based sources. This off-axis OM geometry, implemented via custom in-vacuum mirrors with an integrated beam-output aperture, eliminates pushing-beam light shifts and suppresses fluorescence leakage, thereby minimizing decoherence. Light shifts and decoherence are further evaluated using continuous Raman-Ramsey interferometry. Compactness is achieved through permanent magnets and an integrated opto-mechanical design. Compared with previous efforts employing two-stage cooling to realize a three-dimensionally cooled atomic beam \cite{Black2020,Kwolek2020,Kwolek2022}, our single-cell source achieves full three-dimensional cooling within a single spatial region and features intrinsically ultra-low light shifts. These advances establish a pathway toward practical, high-performance continuous cold-atom beam sources suited for field-deployable interferometers and clocks.

\section{Principle}
\label{sec:theory}

\begin{figure}[t]
\includegraphics[width=0.9\linewidth]{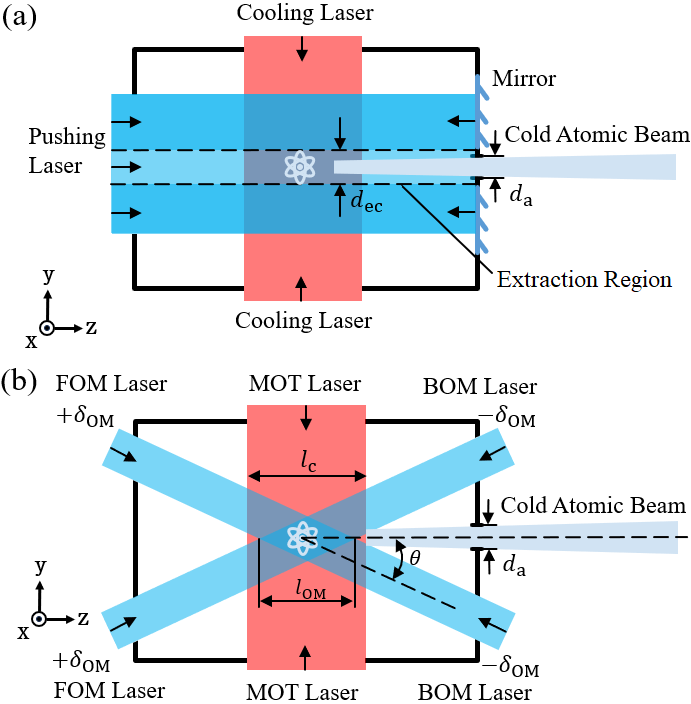}
\caption{\label{fig:principlecompara}
Principles of (a) a classical MOT-based single-cell beam source and (b) an off-axis moving-OM single-cell source. The MOT laser beams counterpropagate along $x$ and $y$ axes, with the $x$-axis beams omitted omitted from the schematic for clarity. The Forward OM (FOM) and backward OM (BOM) beams are oriented at angle $\theta$ relative to the extraction axis. $d_{\rm ec}$ is the diameter of the extraction region defined by the unidirectional component of the pushing laser beam; $d_{\rm a}$ is the diameter of the mechanical output aperture; $l_{\rm c}$ is the cooling length defined by the MOT laser beams; $l_{\rm OM}$ is the OM interaction length defined by the OM laser beams.}
\end{figure}

Classical MOT-based single-cell beam sources rely on extraction regions formed by apertures in the pushing-beam mirrors, as shown in Fig.~\ref{fig:principlecompara}(a). Within the extraction region, atoms are driven out by a unidirectional pushing beam. However, atoms in this region are heated rather than cooled along the longitudinal direction, typically resulting in longitudinal temperatures of several tens of millikelvin \cite{Lu1996}.

To eliminate this heating while maintaining efficient extraction, we integrate an off-axis moving optical molasses (OM) with a 2D MOT, as illustrated in Fig.~\ref{fig:principlecompara}(b). The 2D MOT confines atoms within a cylindrical region and provides transverse cooling, while the moving OM-generated by symmetrically shifting the forward (FOM) and backward (BOM) OM beam frequencies by $\pm\delta_{\mathrm{OM}}$-cools atoms along the extraction axis and simultaneously tunes their mean velocity. The shifted OM acts in a moving reference frame defined by two pairs of counterpropagating beams intersecting at an angle $\theta$ relative to the extraction axis. After sufficient cooling, the mean atomic velocity is given by
\begin{equation}
 v = \frac{\lambda}{\cos\theta}\,\delta_{\rm OM},
\label{eq:v_set}
\end{equation}
where $\lambda$ is the optical wavelength.

\begin{figure}[t]
\includegraphics[width=0.90\linewidth]{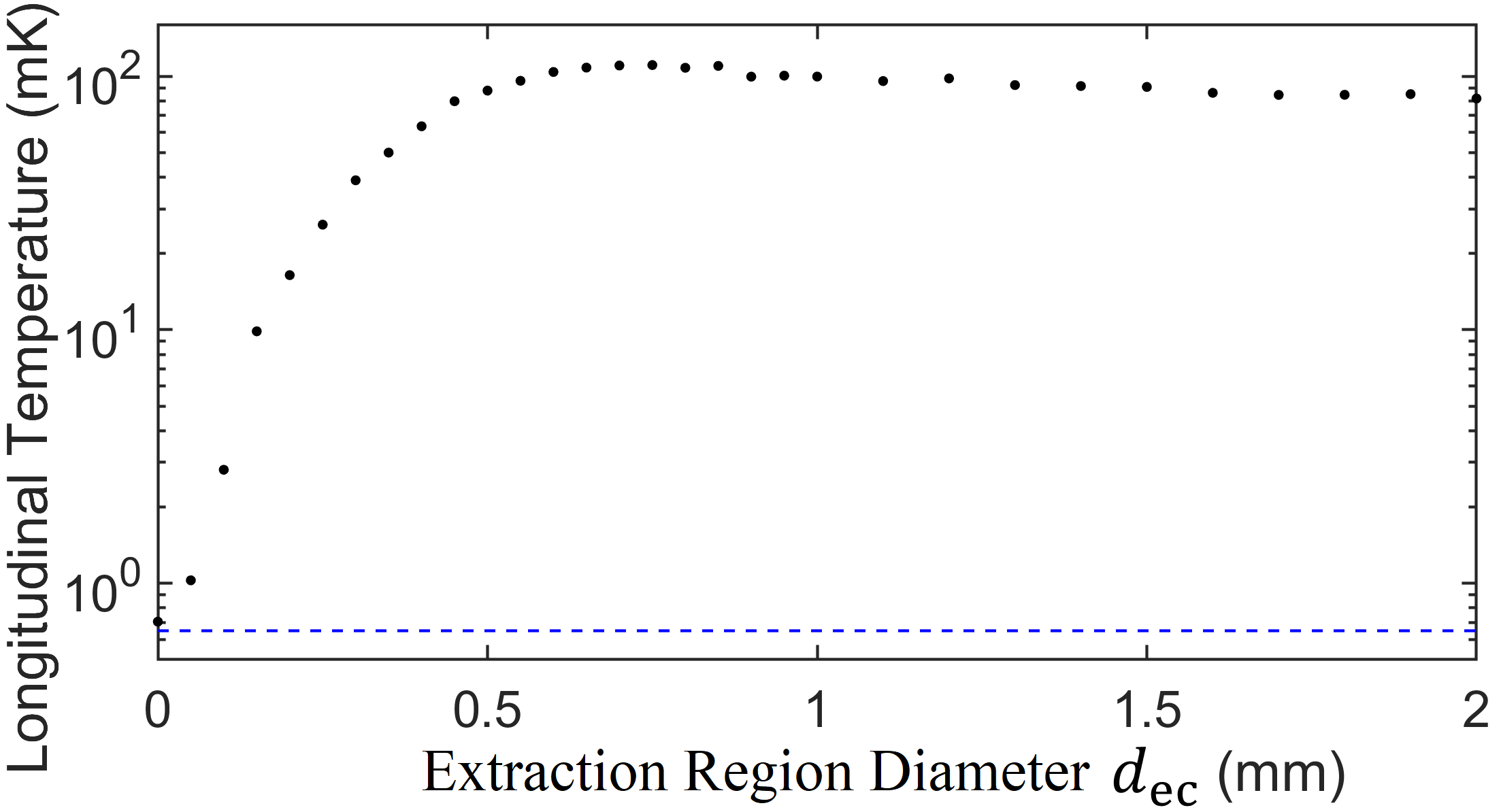}
\caption{\label{fig:on_Extraction_Column}
Simulated longitudinal temperature of a 2D$^+$ MOT beam as a function of extraction region diameter $d_{\rm ec}$, defined by the unidirectional pushing laser beam. The blue dashed line indicates the simulated longitudinal temperature for the off-axis OM configuration.}
\end{figure}

We developed a two-level Doppler cooling and trapping model to evaluate the longitudinal heating in different single-cell configurations. Atoms are treated as massive particles subjected to stochastic scattering forces from the cooling lasers, including both fluctuating photon absorption and isotropic spontaneous emission. The average scattering rate $R_{{\rm scatt},i}$ and corresponding force $\boldsymbol{F}_{{\rm scatt},i}$ for each laser beam $i$ are
\begin{equation}
\begin{aligned}
 R_{{\rm scatt},i} &= \frac{\Gamma}{2}\,
 \frac{s_i}{1 + \sum_{i=1}^{n}s_i + (2\delta_i/\Gamma)^2}, \\
 \boldsymbol{F}_{{\rm scatt},i} &= \hbar \boldsymbol{k}_i\, R_{{\rm scatt},i},
\label{eq:scattering_force}
\end{aligned}
\end{equation}
where $\Gamma$ is the natural linewidth, $s_i = I_i/I_{\mathrm{sat}}$ is the saturation parameter of beam $i$ (with $I_i$ the local laser intensity and $I_{\mathrm{sat}}$ the saturation intensity), $\delta_i$ the detuning including Doppler and Zeeman shifts, and $\boldsymbol{k}_i$ the wavevector. Photon-absorption fluctuations follow Poisson statistics with standard deviation $\sqrt{N_i}$, where $N_i = R_{{\rm scatt},i} t$ is the mean number of absorbed photons during time $t$. Spontaneous emission is modeled as an isotropic random walk with $\sum_i N_i$ steps in momentum space.

Fig.~\ref{fig:on_Extraction_Column} shows simulated longitudinal temperatures of a 2D$^+$ MOT beam as a function of extraction region diameter $d_{\rm ec}$. In the model, $d_{\rm ec}$ is varied to evaluate the heating contribution of the unidirectional pushing beam. The aperture diameter $d_{\rm a}$ is kept constant in the simulation, although $d_{\rm ec}$ is effectively determined by $d_{\rm a}$ in most experiments. The simulation shows that reducing $d_{\rm ec}$ to zero lowers the longitudinal temperature by nearly two orders of magnitude, confirming that the extraction region is a major source of heating in single-cell MOT configurations. For comparison, the longitudinal temperature is simulated for the off-axis OM configuration [Fig.~\ref{fig:principlecompara}(b)], which is equal to that of the classical MOT configuration with $d_{\rm ec} = 0$. This indicates that the off-axis moving OM configuration operates without heating from the unidirectional pushing region. This enables the source to provide significantly more effective longitudinal cooling compared with the other single-cell sources.

\begin{figure}[t]
\includegraphics[width=0.90\linewidth]{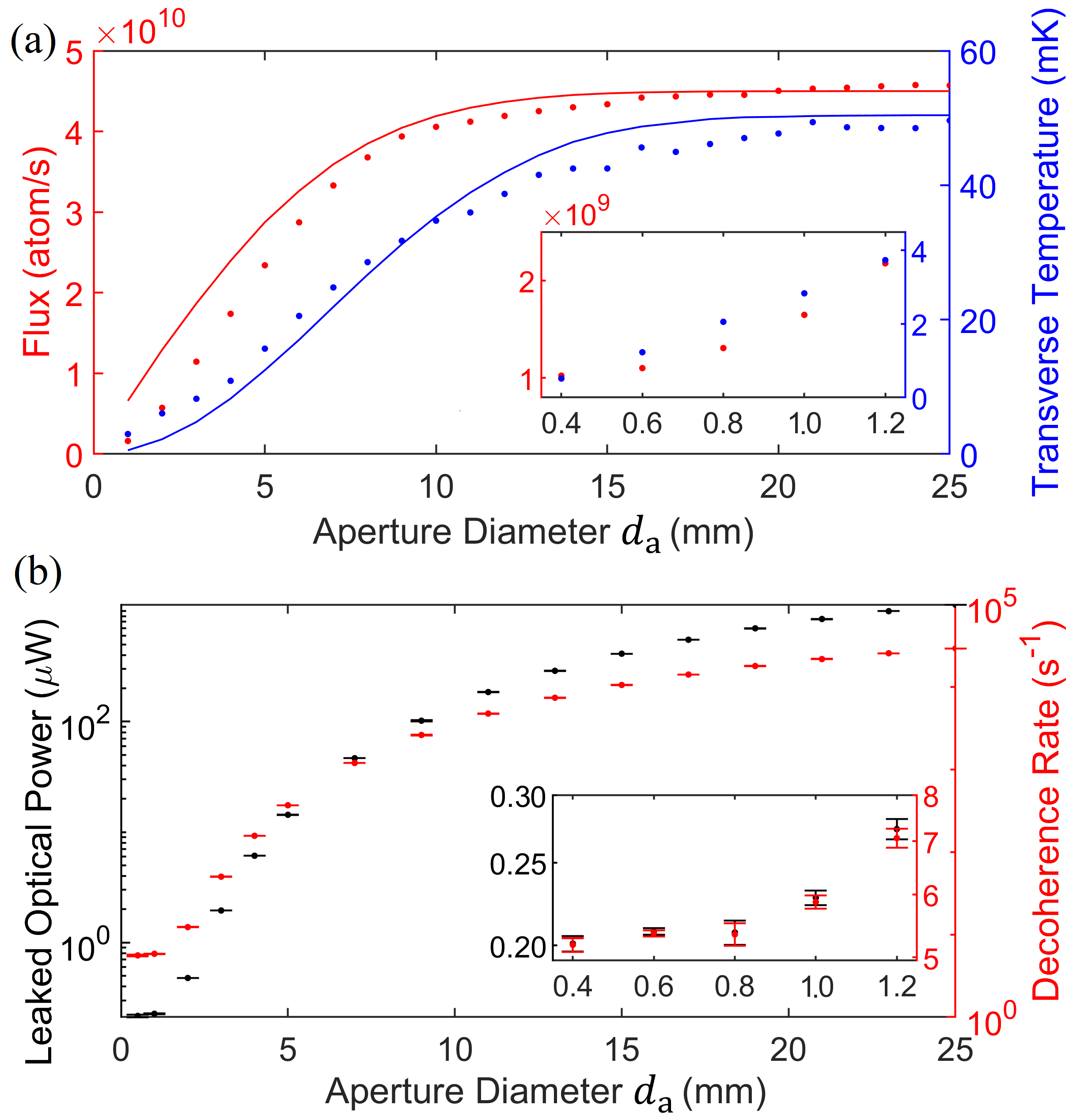}
\caption{\label{fig:On_Hole_Diameter}
Effect of output aperture size $d_{\rm a}$. (a) Theoretical atomic flux and transverse temperature as functions of $d_{\mathrm{a}}$. Dots: numerical simulations. Lines: analytical estimates from Eq.~\ref{eq:alpha}, assuming $l_{\rm c}=50~\mathrm{mm}$ and mean velocity $11.5~\mathrm{m/s}$. (b) Predicted near-resonant optical power leaked into the downstream cell from ray-tracing simulations, and the corresponding decoherence rate estimated from the scattering rate.}
\end{figure}

The transverse temperature of the off-axis OM $+$ 2D MOT configuration is governed by the cooling length $l_{\rm c}$ and the output aperture diameter $d_{\rm a}$, which define a maximum collimation angle
\begin{equation}
 \alpha = \frac{d_{\mathrm{a}}}{l_{\mathrm{c}}}.
\label{eq:alpha}
\end{equation}
The transverse temperature and atomic flux depend on the aperture diameter $d_{\mathrm{a}}$, as the aperture provides transverse collimation by spatially selecting the sufficiently cooled, low-divergence portion of the atomic beam. The simulated flux and transverse temperature as functions of $d_{\rm a}$ are shown in Fig.~\ref{fig:On_Hole_Diameter}(a). For a fixed $l_{\rm c}$, decreasing $d_{\rm a}$ reduces the transverse temperature but also decreases atomic flux, illustrating a trade-off between flux and temperature. The numerical results agree well with the analytical relation of Eq.~\ref{eq:alpha}.

During the cooling process, near-resonant light is scattered in random directions. The output aperture constitutes the only optical path for this scattered light to escape the cooling cell. Consequently, reducing the aperture diameter decreases the leaked optical power and the resulting decoherence. The leaked power is quantified using ray-tracing simulations that treat both the atomic cloud and surrounding structures as secondary light sources. Details are provided in Appendix~\ref{app:simulation}. This leaked light drives spontaneous scattering that leads to decoherence, defined as loss of phase coherence in the atomic superposition state. By assuming that each scattering event leads to decoherence, the decoherence rate is taken to be equal to the scattering rate in Eq.~\ref{eq:scattering_force}, with the leaked light assumed to be isotropically polarized. A more accurate estimation of scattering-induced decoherence would require a detailed quantum treatment of the scattering processes \cite{Moore2023}. Reducing $d_{\rm a}$ suppresses both quantities of light and decoherence. A $0.8~\mathrm{mm}$ aperture isolates most stray light and fluorescence, limiting the decoherence rate to $5.34~\mathrm{s^{-1}}$ while maintaining a low transverse temperature of $62.8~\mu\mathrm{K}$ and a flux of $6.1\times10^9~\mathrm{atoms/s}$ (calculated from Eq.~\ref{eq:alpha}), at a mean velocity of $11.5~\mathrm{m/s}$ and a cooling length of $l_{\rm c}=50~\mathrm{mm}$. In the meantime, the longitudinal temperature provided by the off-axis OM is simulated to be $650~\mu\mathrm{K}$.

Overall, the model demonstrates that eliminating the unidirectional pushing beam and employing an off-axis moving-molasses geometry transforms the single-cell MOT from a heating-limited system into one that achieves complete three-dimensional cooling with tunable beam velocity and intrinsically low decoherence-establishing the physical foundation of our continuous cold-atom beam source.

%---------------------------------------------------------------------------------------%

\section{Apparatus}

\begin{figure*}[t]
\includegraphics[width=0.95\linewidth]{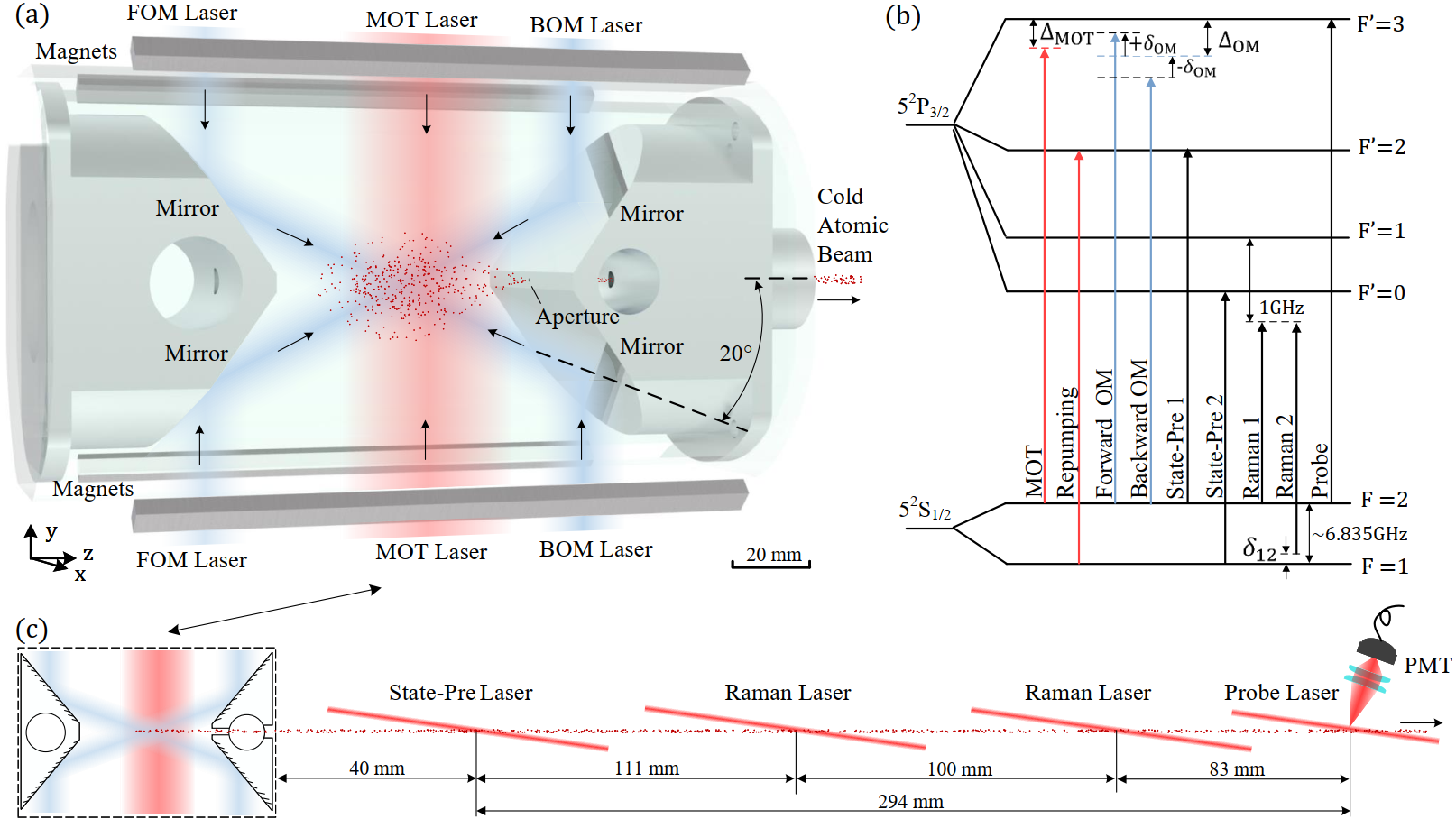}
\caption{\label{fig:Apparatus}
Apparatus schematic. (a) Vacuum cell containing the 2D MOT and the off-axis moving OM. Directions of the laser beams are indicated by the arrows. Four mirrors define the optical paths for the forward OM (FOM) and backward OM (BOM) laser beams (blue). The MOT laser beams counterpropagate along the $x$ and $y$ axes, with the $x$-axis beams omitted from the schematic for clarity. Both the cooling length $l_{\rm c}$ and OM interaction length $l_{\rm OM}$ are about $50~\mathrm{mm}$. The cell is illustrated as transparent for clarity. (b) Energy-level diagram and laser frequencies. $\Delta_{\rm MOT}$ and $\Delta_{\rm OM}$ denote MOT and OM detunings; $\delta_{\rm OM}$ is the OM frequency shift; $\delta_{12}$ is the two-photon detuning of Raman lasers. (c) Locations of laser beams used to characterize the cold-atom beam (not to scale).}
\end{figure*}

As shown in Fig.~\ref{fig:Apparatus}(a), the $^{87}$Rb source integrates an off-axis moving OM with a 2D MOT. The OM region overlaps the 2D MOT, providing both transverse and longitudinal cooling within the same cell. 

The 2D MOT is formed by two orthogonal pairs of counterpropagating, circularly polarized laser beams, detuned by $\Delta_{\rm MOT}=-4\Gamma$ from the $F=2\to F'=3$ cycling transition of the D$_2$ line, where $\Gamma=2\pi\times6~\mathrm{MHz}$ is the natural linewidth. Repumping laser light for the $F=1\to F'=2$ transition is generated via phase modulation of the MOT laser using an electro-optic modulator (EOM). The relevant $^{87}$Rb level structure and laser frequencies are shown in Fig.~\ref{fig:Apparatus}(b). A cylindrical quadrupole magnetic field is used in conjunction with the MOT laser beams to trap atoms. The magnetic field gradient along the $x$ and $y$ directions is $10~{\rm G/cm}$ approximately.

The off-axis OM consists of two pairs of counterpropagating beams, oriented at $20^{\circ}$ relative to the atomic output axis and arranged in a lin$\perp$lin polarization configuration to optimize longitudinal cooling. The OM laser detuning is set to $\Delta_{\rm OM}=-5\Gamma$, chosen to minimize the longitudinal temperature. To control the mean velocity, the FOM and BOM lasers are symmetrically shifted by $\pm\delta_{\rm OM}$, cooling atoms into the moving frame.

Under regular operating conditions, the saturation parameter of the MOT laser beams is set to $s = 7.19$, with an approximately $50~{\rm mm}\times 25~{\rm mm}$ beam profile (saturation intensity $I_{\rm sat} = 1.67~{\rm mW/cm^2}$). The repumping laser light operates with a saturation parameter of $s = 0.72$.

The off-axis OM laser beams have a saturation parameter of $s = 3.53$ and an approximate beam diameter of $18~{\rm mm}$. The moving OM frequency detunings are set to $\pm 10~{\rm MHz}$, corresponding to a mean atomic velocity of $11.5~{\rm m/s}$ under regular operating conditions.

Compactness of the source is achieved with the optical and magnetic design. In-vacuum mirrors reflect the OM laser beams at $20^{\circ}$ relative to the atomic output axis and direct light transmitted from the cooling region out of the cell, thereby reducing near-resonant light into the downstream cell. The mirrors are created by directly polishing the inner surfaces of two custom-designed aluminum blocks, providing stable alignment that enhances both compactness and robustness. An aperture with $d_{\rm a}=0.8~\mathrm{mm}$ on the right side defines the atomic beam output. The magnetic field of the 2D MOT is generated by four permanent magnets, and the zero intensity line of the magnetic field can be fine-tuned with kinematic mounts to maximize the output atomic flux.

The MOT and OM beams have effective sizes of $50~\mathrm{mm}\times 25~\mathrm{mm}$ and $18.5~\mathrm{mm}$ in diameter, respectively, defining a cooling length $l_{\mathrm{c}} = 50~\mathrm{mm}$ and an OM interaction length $l_{\mathrm{OM}} \approx 50~\mathrm{mm}$. The overall length of the apparatus, including the vacuum structure, is approximately $170~\mathrm{mm}$. These compact dimensions enable full three-dimensional cooling and stable atomic-beam extraction within a single vacuum cell.

\section{Characterization}
\subsection{Temperature, Velocity, and Flux}

\begin{figure}[t]
\includegraphics[width=0.8\linewidth]{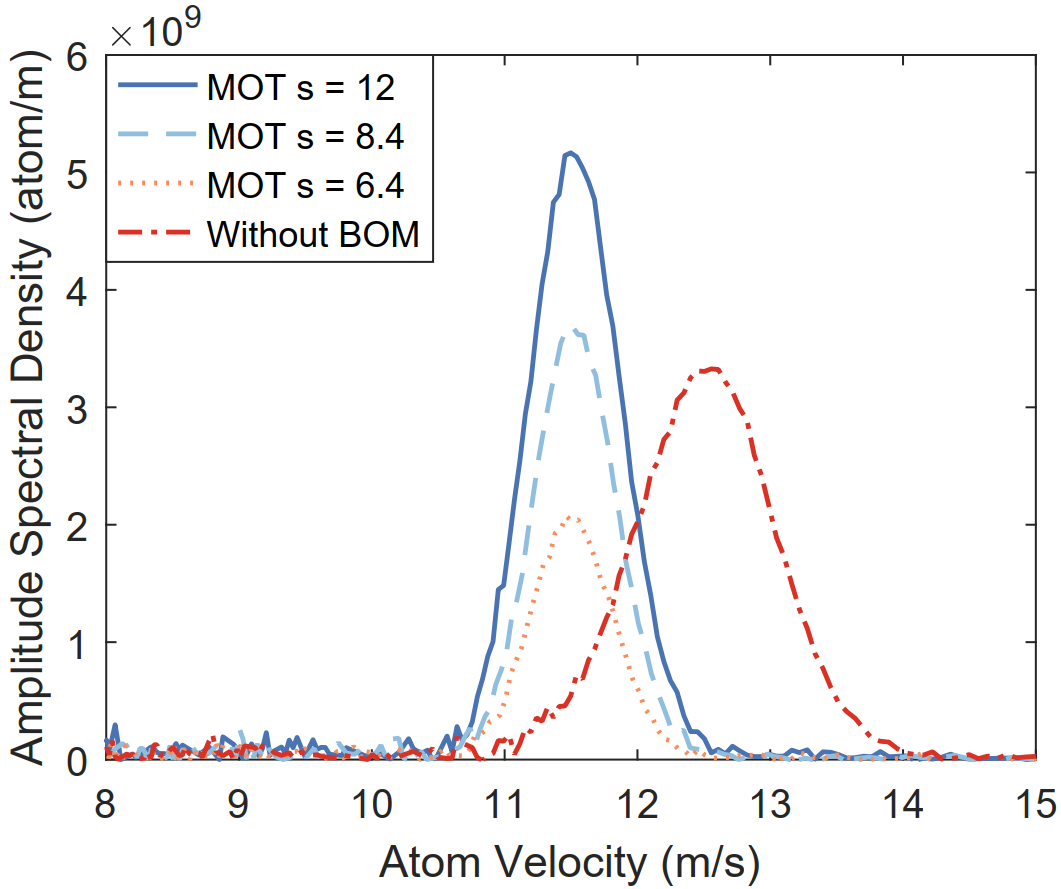}
\caption{\label{fig:Longti_dis}
Longitudinal velocity distribution measured over a $294~\mathrm{mm}$ flight distance, for different saturation parameters $s = I/I_{\rm sat}$ of the transverse MOT lasers (where $I_{\rm sat} = 1.67 ~{\rm mW/cm^2}$). The distribution obtained without the BOM is also shown at $s = 12$.}
\end{figure}

\begin{figure}[t]
\includegraphics[width=0.8\linewidth]{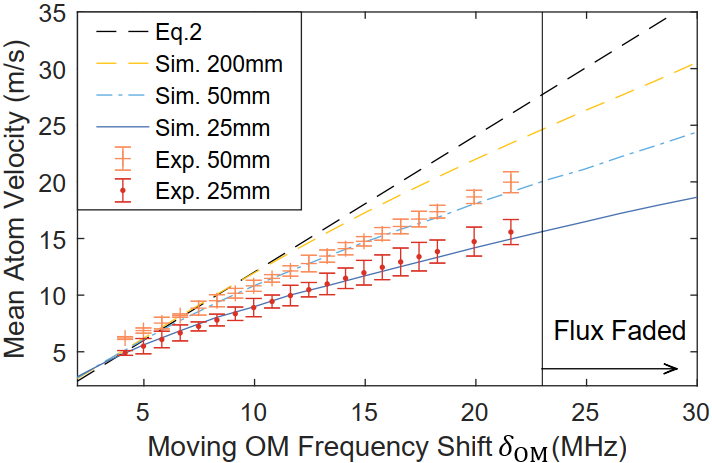}
\caption{\label{fig:Velocity_sweep_L}
Mean atom velocity as a function of the moving-OM frequency shift $\delta_{\rm OM}$ for $l_{\rm OM}=50~\mathrm{mm}$ and $25~\mathrm{mm}$. Experimental data ("Exp.") are compared with simulations ("Sim."). The $25~\mathrm{mm}$ length is realized by partially blocking the OM beams. A theoretical slope of $1.2~(\mathrm{m/s})/\mathrm{MHz}$ from Eq.~\ref{eq:v_set} is shown for reference.}
\end{figure}

Atoms extracted from the source are initially in the $F=2$ hyperfine state. The atomic beam flux is measured via fluorescence detection. A probe laser beam, resonant with the $F=2\to F'=3$ transition with an intensity of $1.38~{\rm mW/cm^2}$ (saturation intensity $I_{\rm sat} = 1.67~{\rm mW/cm^2}$), propagates perpendicular to the atomic beam and induces fluorescence proportional to the number of atoms passing through the probe region. The fluorescence is collected by an optical system and detected by a photomultiplier tube (PMT), from which the atomic flux in the $F=2$ state is inferred. The quantization axis is defined along the probe-beam direction, and the probe beam is retro-reflected with $\sigma+/\sigma-$ polarization.

A state-preparation laser beam, with frequencies labeled State-Pre~1 and State-Pre~2 in Fig.~\ref{fig:Apparatus}(b), is positioned between the atomic source and the probe beam. When switched on, the state-preparation laser optically pumps atoms into the $F=1$ hyperfine state, which is dark to the probe beam. When switched off, atoms remain in the $F=2$ state and can be detected by fluorescence.

The longitudinal temperature and mean velocity are measured using a time-of-flight (TOF) method \cite{Dieckmann1998,Catani2006}. Upon rapid switching off of the state-preparation laser, faster atoms in the $F=2$ state reach the probe region earlier than slower ones, producing a time-resolved fluorescence signal from which the longitudinal velocity distribution is derived.

\begin{figure}[t]
\includegraphics[width=0.90\linewidth]{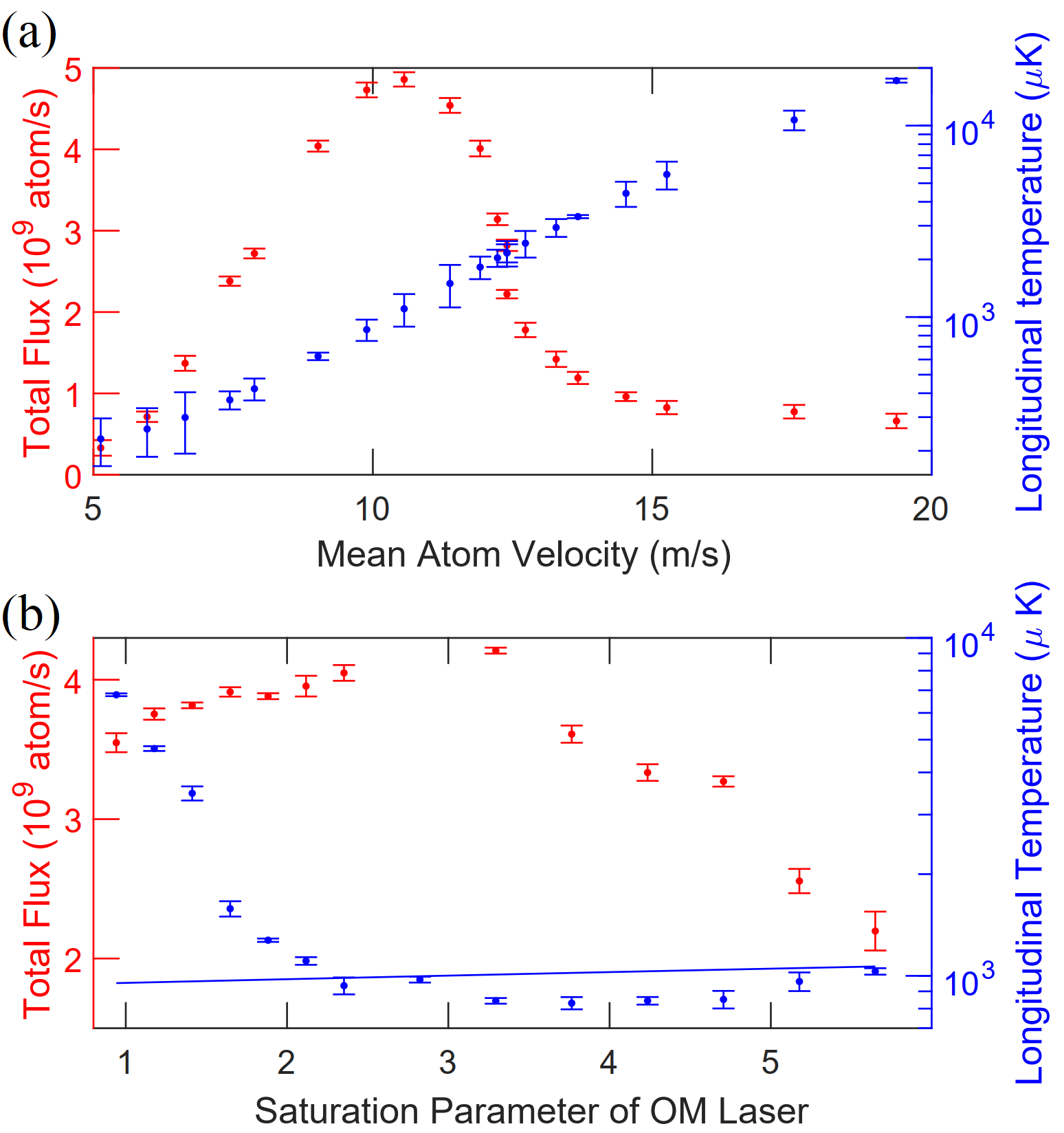}
\caption{\label{fig:longTempFluxon}
Atomic flux and longitudinal temperature versus mean velocity (a) and OM saturation parameters $I/I_{\mathrm{sat}}$ (b). $I_{\rm sat} = 1.67 ~{\rm mW/cm^2}$. The blue curve in (b) shows temperatures predicted by the Doppler cooling theory \cite{Lett89}.}
\end{figure}

The state-preparation and probe laser beams are shaped to a $1~\mathrm{mm}$ thickness along propagation of the atomic beam. For a $294~\mathrm{mm}$ flight length and a mean velocity of $10~\mathrm{m/s}$, the velocity resolution is $0.05~\mathrm{m/s}$, corresponding to $5~\mu\mathrm{K}$ in temperature.

Fig.~\ref{fig:Longti_dis} shows a representative longitudinal velocity distribution. The longitudinal temperature is extracted from the FWHM of the distribution, while the flux is determined from the integrated spectral density. In the example shown, the temperature is $800\pm200~\mu\mathrm{K}$ at a mean velocity of $11.5~\mathrm{m/s}$, corresponding to a flux of $4.9\times10^9~\mathrm{atoms/s}$. Increasing MOT power raises the flux but affects the temperature only weakly. When the BOM laser is disabled, the velocity distribution broadens and the mean velocity increases, indicating that the laser provides deceleration and cooling.

\begin{figure}[t]
\includegraphics[width=0.9\linewidth]{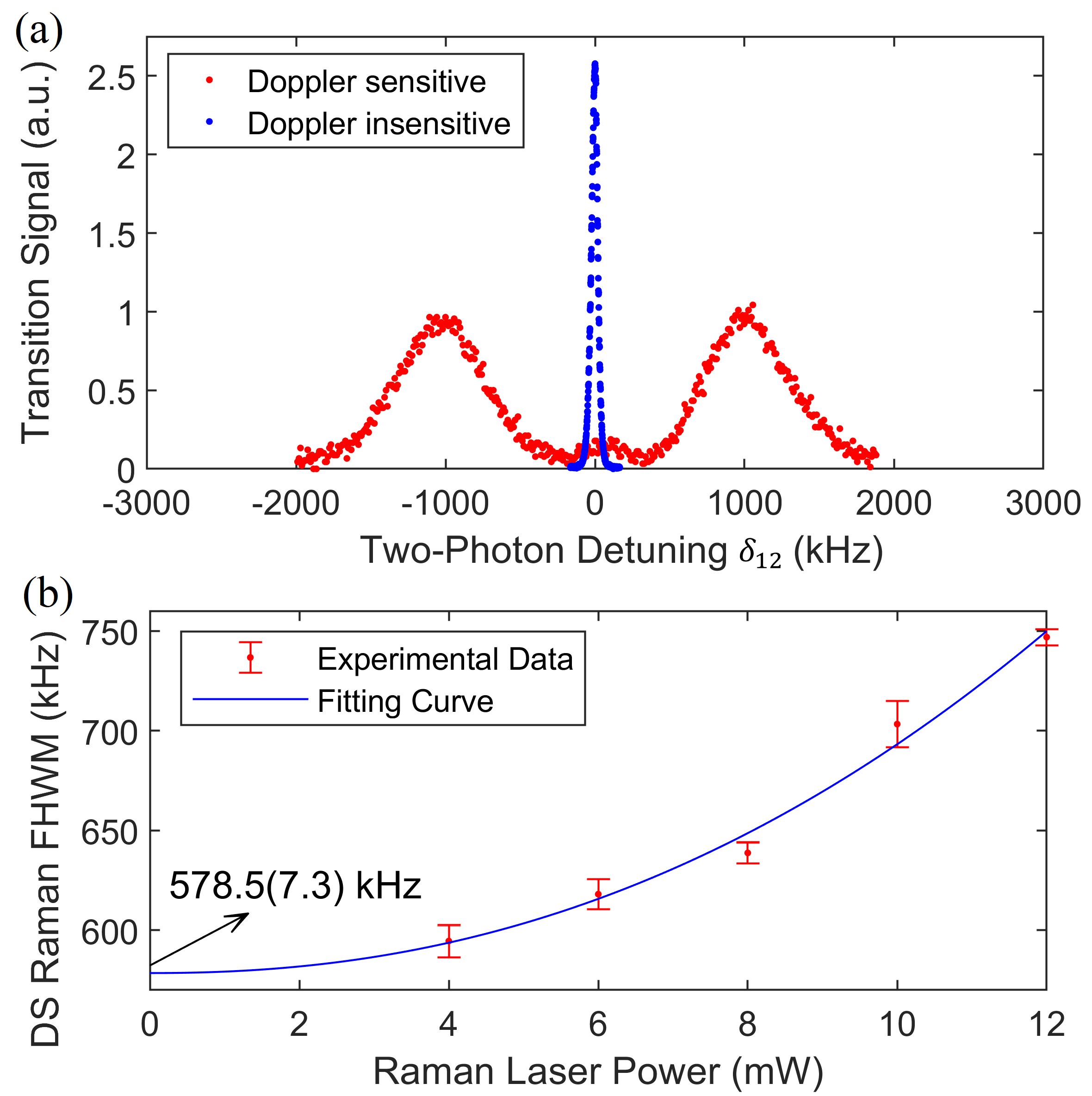}
\caption{\label{fig:MeansurementRaman}
(a) Raman transition spectrum at a Rabi phase of $\pi$. (b) Doppler-sensitive Raman linewidth (FWHM) versus Raman power. Quadratic fitting accounts for spectral leakage. Signal amplitude in (a) is not to scale.}
\end{figure}

The polished aluminum mirrors used for the OM laser beams exhibit a reflectivity of approximately $91\%$ over a broad wavelength range from $450$ to $2000~{\rm nm}$. As metallic mirrors, they modify the polarization of the reflected OM beams, which affects the longitudinal cooling efficiency. Without polarization optimization, the longitudinal temperature is typically in the $1.2\text{-}2.0~{\rm mK}$. By optimizing the OM beam polarization, the longitudinal temperature can be reduced to $0.2\text{-}0.8~{\rm mK}$, depending on the atomic velocity.

When the moving OM extracts atoms from zero mean velocity, the cooling time may be insufficient for atoms to reach the moving-frame velocity predicted by Eq.~\ref{eq:v_set}. Fig.~\ref{fig:Velocity_sweep_L} shows that the measured mean velocity increasingly deviates from the theoretical slope (black dashed line calculated from Eq.~\ref{eq:v_set}, corresponding to an infinite OM interaction length) as $\delta_{\rm OM}$ grows. Simulations show that longer OM interaction lengths $l_{\rm OM}$ yield better agreement to the theoretical slope, indicating that the achievable velocity is fundamentally limited by $l_{\rm OM}$.

The dependence of flux and temperature on mean velocity and OM saturation parameter is summarized in Fig.~\ref{fig:longTempFluxon}, where the mean velocity is tuned via $\delta_{\rm OM}$. Fig.~\ref{fig:longTempFluxon}(a) shows that at mean velocities above $11~\mathrm{m/s}$, the temperature rises and the flux decreases because atoms across the cooling region too quickly. The lowest longitudinal temperature experimentally achieved in our source is $231(65)~\mu\mathrm{K}$ at a mean atomic velocity of $5.1~{\rm m/s}$.
However, the flux also decreases at lower velocities. TOF measurements at $93$ and $294~\mathrm{mm}$ flight distances show no significant difference, indicating that the reduced flux is not caused by the atomic divergence. The observed reduction in atomic flux at lower mean velocities can be attributed to the increasingly stringent transverse velocity selection imposed by the output aperture. As the mean atomic velocity $v$ decreases, the accepted transverse velocity range scales as $\alpha \cdot v$, where $\alpha$ is the collimation angle determined by the aperture geometry Eq.~\ref{eq:alpha}. Consequently, a lower mean velocity corresponds to a narrower transverse acceptance window, resulting in fewer atoms being transmitted through the aperture and thus a reduced atomic flux.

Fig.~\ref{fig:longTempFluxon}(b) shows that flux decreases as saturation parameter of OM laser increases, even with constant MOT intensity. This behavior arises from the saturation of scattering forces during multi-dimensional cooling in the same region. From Eq.~\ref{eq:scattering_force}, raising the OM saturation parameter from 1 to 6 reduces the transverse MOT scattering force from $3.76\times10^{-21}~\mathrm{N}$ to $2.69\times10^{-21}~\mathrm{N}$. Because the MOT along $x$ and $y$ jointly confines the atoms, the squared reduction of this force corresponds to an approximately $1.95$-fold drop in trapping efficiency, consistent with the observed twofold reduction in flux. The ensemble in the cooling region thus behaves analogously to a balloon in momentum space: compression along one dimension leads to expansion along the others.

At OM saturation parameters below $s = 3$, equilibrium between cooling and heating is not reached, leading to higher longitudinal temperatures. Above this point, the temperature increases only slowly, in qualitative agreement with Doppler theory \cite{Lett89}, though the measured values are somewhat lower.

\begin{figure}[t]
\includegraphics[width=0.85\linewidth]{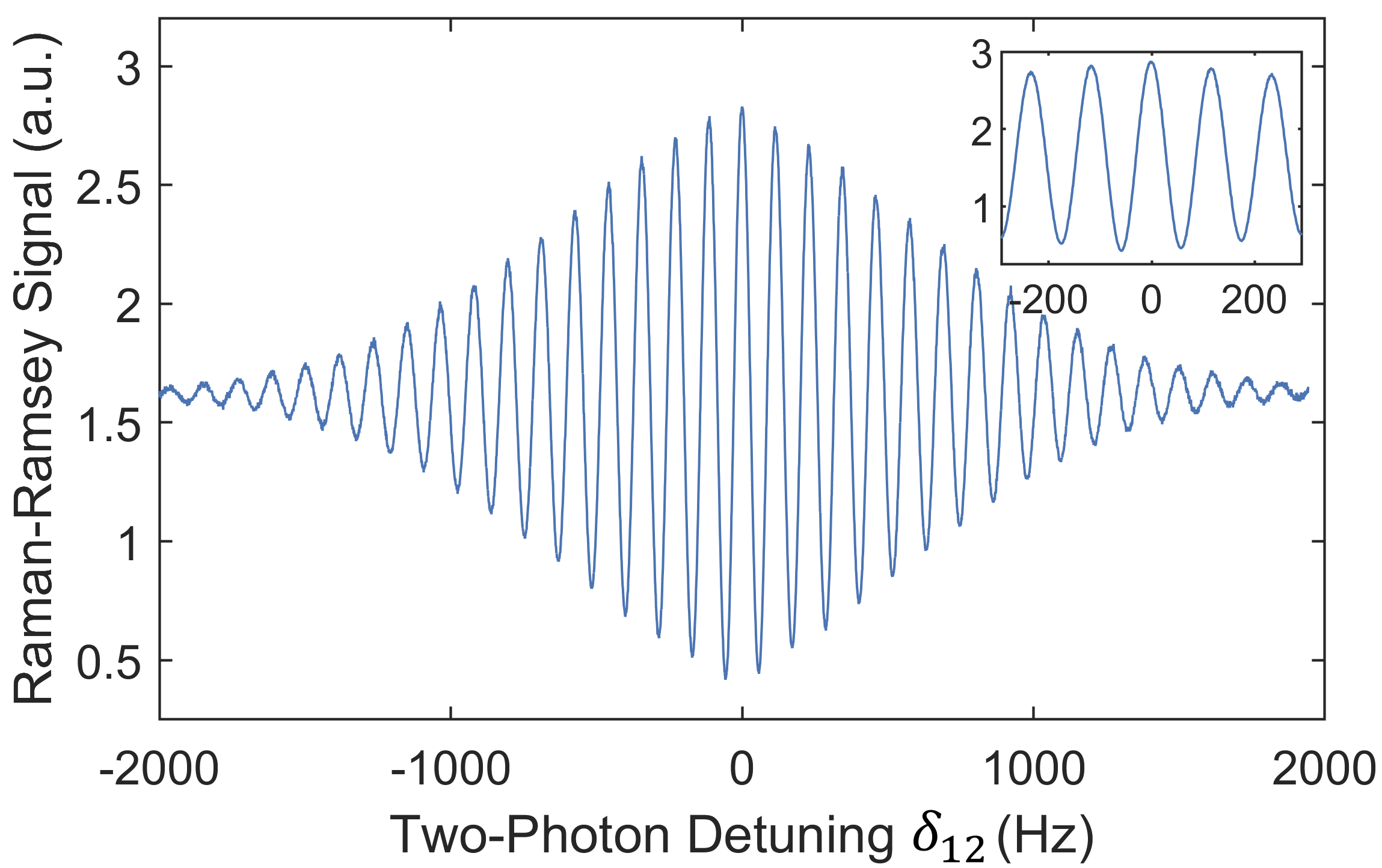}
\caption{\label{fig:Ramsey_demonstration}
Continuous spatial-domain Raman-Ramsey interference with interrogation length $L=100~\mathrm{mm}$ and mean atomic-beam velocity $v=11.5~\mathrm{m/s}$. Inset: central fringe (five periods).}
\end{figure}

\begin{figure*}[t]
\includegraphics[width=0.9\linewidth]{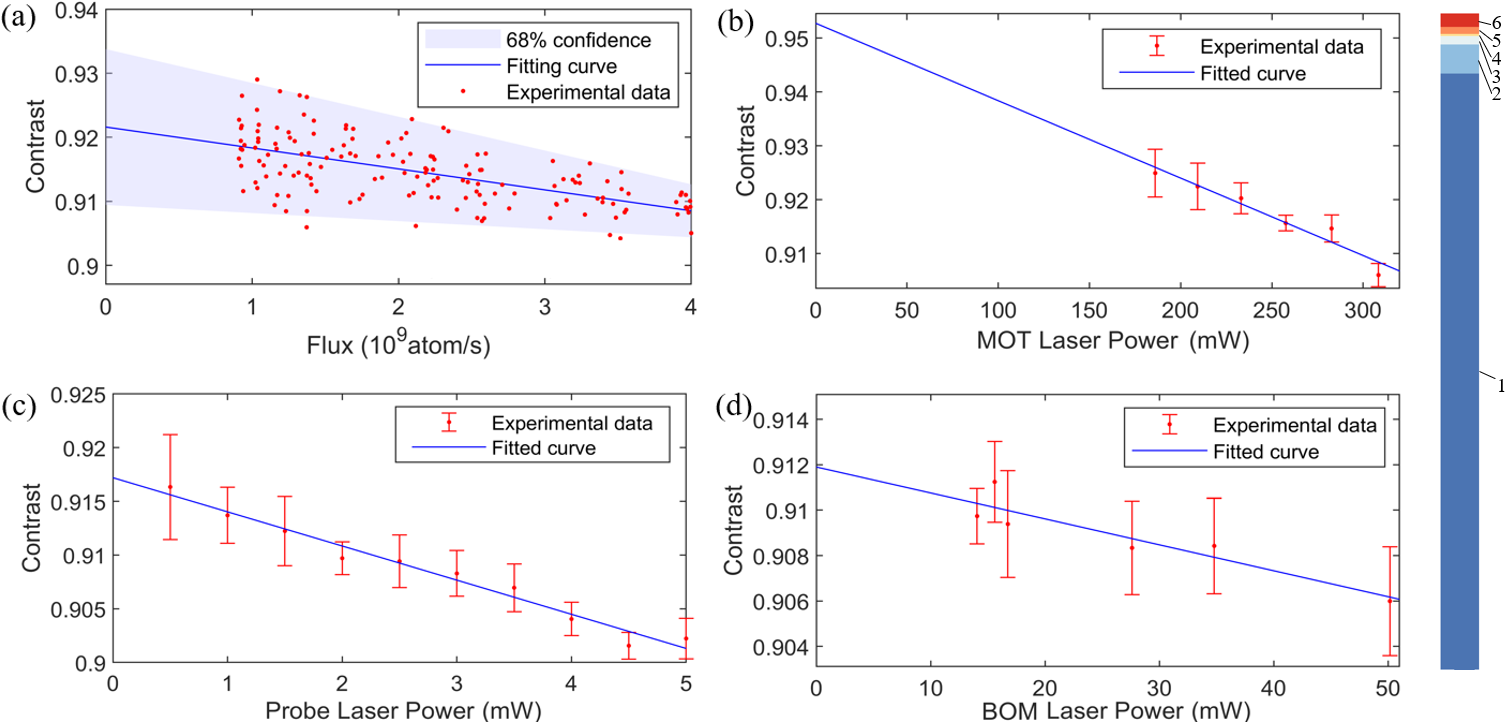}
\caption{\label{fig:Alldecoherencefactor}
Continuous Raman-Ramsey fringe contrast as functions of (a) atomic flux, (b) MOT power, (c) probe power, and (d) BOM power. Increasing atomic flux in (a) reveals fluorescence-induced decoherence. The dependences on MOT and BOM power in (b) and (d) primarily reflect the effect of stray light. Because variations in optical power also affect the atomic flux, the data in (b) and (d) are corrected using the fit obtained from (a). The contribution from the FOM beams is negligible, while probe-induced loss originates outside the atomic source region. Right: Relative contributions to the reduction of the Ramsey contrast under typical operating conditions. The bars represent the relative contributions of different mechanisms rather than absolute contrast values: (1) Ramsey contrast, $90.85\%$ (dark blue); (2) MOT stray light, $4.43(30)\%$ (sky blue); (3) atomic fluorescence, $1.28(82)\%$ (light cyan); (4) BOM stray light, $0.33(20)\%$ (gold); (5) probe stray light/fluorescence, $1.06(24)\%$ (orange); and (6) unidentified contributions, $2.05\%$. Experimental parameters: $L=100~\mathrm{mm}$, $v=11.5~\mathrm{m/s}$.}
\end{figure*}

The transverse velocity distribution is measured using Doppler-sensitive stimulated Raman transitions \cite{Kasevich1991,Moler1992}. Raman~1, detuned $1~\mathrm{GHz}$ below the D$_2$ line, is phase-modulated with an EOM near $6.835~\mathrm{GHz}$ to generate Raman~2 frequency. The modulation depth is adjusted to cancel light shifts from the Raman lasers. Raman power is tunable to suppress spectral broadening, and the beam is shaped to yield a $50~\mathrm{kHz}$ transit-time linewidth, corresponding to temperature resolution of $0.7~\mu\mathrm{K}$. To drive Doppler-sensitive transitions, the beam is retroreflected with lin$\perp$lin polarization.

The Doppler-sensitive Raman spectrum is a convolution of the Doppler-insensitive spectrum (set by transit time) and Doppler broadening from transverse velocities [Fig.~\ref{fig:MeansurementRaman}(a)]. A spectral leakage in Doppler-insensitive Raman transition, induced by the rectangular temporal profile of the Raman beam shapes, broadens the linewidth of the spectrum. Linewidths are thereby measured at multiple Raman laser powers and extrapolated to zero. The intercept in Fig.~\ref{fig:MeansurementRaman}(b) yields a linewidth of $578.5(7.3)~\mathrm{kHz}$, corresponding to a transverse temperature of $94(5)~\mu\mathrm{K}$ ($62.8~\mu\mathrm{K}$ predicted in Section~\ref{sec:theory}). Notably, the transverse temperature is largely insensitive to MOT and OM optical parameters.

Compared with previous single-cell cold-atom beams \cite{Dieckmann1998,Arlt1998,Camposeo2001,Schoser2002,Kohel2003,Xie2022,Park2012,Wang2024,Kellogg2012,Wang2003,Wang2011}, our source achieves a substantially reduced longitudinal temperature as low as $231(65)~\mu\mathrm{K}$ with tunable velocity and comparable transverse temperature. This mitigates contrast loss in interferometers under rotation and acceleration, where longitudinal velocity spread dominates \cite{Black2020,Narducci2022}. Integrating the inertial phase over our measured distributions shows that the $1/e$ width of the interference-contrast envelope increases by a factor of $8.6$ relative to a typical 2D$^+$ MOT source \cite{Wang2024}, enhancing dynamic range for open-loop inertial measurements and improving robustness for closed-loop operation \cite{Sato2025}.

\subsection{Light Shift and Decoherence}

We quantify decoherence and light shift using spatial-domain Raman-Ramsey interferometry \cite{Feng2015}. During interferometric operation, the state-preparation laser operates continuously to prepare the atomic population prior to the Raman interrogation. Two circularly polarized Raman laser beams separated by $L=100~\mathrm{mm}$ implement $\pi/2$ pulses by adjusting the optical intensity. The Raman region is magnetically shielded, with residual inhomogeneity of $\pm300~\mathrm{nT}$. Four current bars generate a uniform bias field to define the quantization axis. A microwave local oscillator (LO) near $6.835~\mathrm{GHz}$ (Keysight N5183B) is phase-locked to an Rb frequency standard (SRS FS725) to provide stable microwave frequency.

Unlike time-domain implementations with fixed interrogation time $T$, here $T=L/v$ depends on atomic velocity. Fig.~\ref{fig:Ramsey_demonstration} shows a typical fringe at $L=100~\mathrm{mm}$ and $v=11.5~\mathrm{m/s}$. The signal envelope FWHM ($1.5~\mathrm{kHz}$) is much narrower than the $50~\mathrm{kHz}$ expected from the nominal pulse duration, due to variations in effective pulse duration across the longitudinal velocity distribution. The measured fringe spacing is $57.70(4)~\mathrm{Hz}$, in good agreement with the $57.5~\mathrm{Hz}$ value inferred from $T$; the small difference arises from uncertainties in $L$ and $v$.

\begin{figure}[t]
\includegraphics[width=0.85\linewidth]{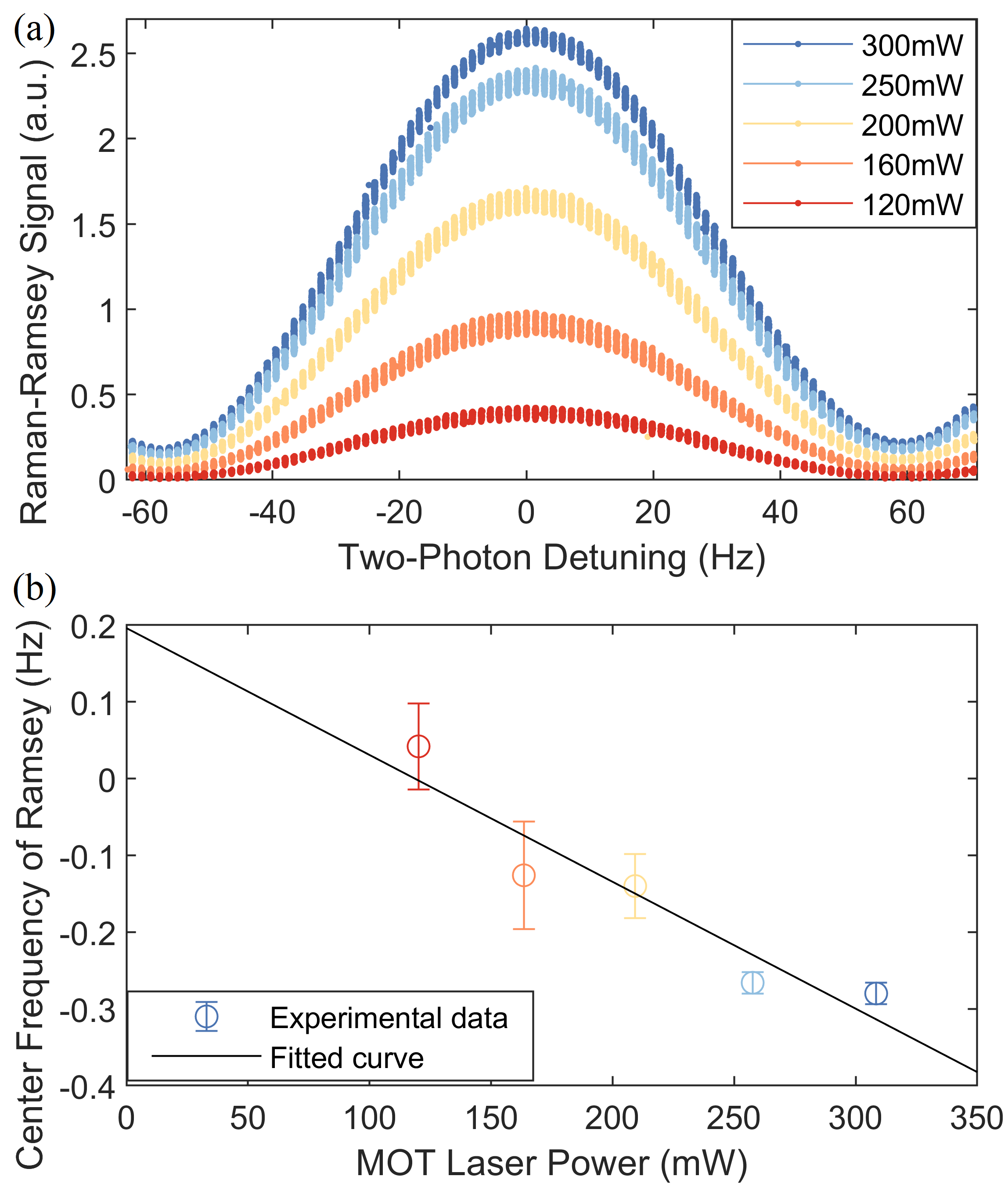}
\caption{\label{fig:dfonMOTP}
(a) Central portion of the Raman-Ramsey fringe at different MOT powers; each trace averages 20 two-photon-detuning sweeps. (b) Center frequency versus MOT power; linear fit gives $-1.653\times10^{-3}~\mathrm{Hz/mW}$. Conditions as in Fig.~\ref{fig:Ramsey_demonstration}. Frequencies are shown relative to $6.834\,680\,907~\mathrm{GHz}$.}
\end{figure}

Decoherence is inferred from the Ramsey-fringe contrast. The contrast is defined as $C = (S_{\rm max}-S_{\rm min})/(S_{\rm max}+S_{\rm min})$, where $S_{\rm max}$ and $S_{\rm min}$ are the maximum and minimum values of the detected Ramsey signal, respectively. Rather than varying $L$ \cite{Kwolek2020}, we sweep atomic flux and laser powers to isolate contrast-loss channels (Fig.~\ref{fig:Alldecoherencefactor}). Flux is tuned via dispenser temperature to reveal fluorescence-induced contrast loss [panel (a)], while MOT and BOM powers are scanned to quantify stray-light effects [panels (b,d)], corrected using the flux dependence in panel (a). The total source-induced contrast loss is $6.04(70)\%$ at $L=100~\mathrm{mm}$, comprising $4.43(30)\%$ from MOT stray light, $0.33(20)\%$ from BOM stray light, and $1.28(82)\%$ from atomic fluorescence. The FOM contribution is negligible, consistent with shielding by the $0.8~\mathrm{mm}$ mirror aperture and off-axis mirror geometry. BOM-induced loss can also be minimized by adjusting the beam direction, though this increases the longitudinal temperature.

In the experiment, the interrogation time is $T=L/v=8.70 ~{\rm ms}$. The decoherence rate is determined from the rate of contrast loss, giving $6.04(70)\%/8.70~{\rm ms} = 6.95(8)~{\rm s^{-1}}$, consistent with the $5.34~{\rm{s}^{-1}}$ predicted by the ray-tracing model (Section~\ref{sec:theory}), validating our fluorescence-isolation design. 

We next evaluate the light shift arising from fluorescence and stray light by measuring the frequency shift of the Ramsey fringes. The interrogation time is modulated (by tuning $v$) to place the LO frequency near center of the fringe, thereby minimizing velocity-to-frequency coupling. Fig.~\ref{fig:dfonMOTP}(a) shows the central portion of the Ramsey fringes at various MOT powers. The vertical spread of the signal reflects amplitude noise due to flux instability rather than LO frequency noise. As the MOT power decreases, the fringe amplitude diminishes because of the reduced flux, while the center frequency of the fringe shifts upward.

\begin{table}[t]
\renewcommand{\arraystretch}{2}
\caption{\label{tab:table1}%
Comparison between our source with representative single-cell continuous cold-atom beam sources reported in the literature. $T_{\rm long}$ and $T_{\rm trans}$ denotes the longitudinal and transverse temperatures, respectively.
}
\begin{ruledtabular}
\begin{tabular}{lcclc}
\textrm{Method}    &
$T_{\rm long}$  &
$T_{\rm trans}$ &
\textrm{Flux}      &
\textrm{Ref.}\\
\colrule
2D MOT             & $11~{\rm K}$\textsuperscript{\ref{footnote1}}   & -  & $6\times10^{10}~{\rm atoms/s}$   & \cite{Schoser2002}          \\
3D MOT             & $14~{\rm mK}$\footnote{Calculated from published FWHM of longitudinal velocity distributions.\label{footnote1}}   & $327~\mu{\rm K}$\footnote{Calculated from the published divergence angles and mean velocities of the atomic beams.\label{footnote2}} & $5\times10^9~{\rm atoms/s}$   & \cite{Lu1996}          \\
$\rm 2D^+MOT$        & $21~{\rm mK}$\textsuperscript{\ref{footnote1}}   & $219~\mu{\rm K}$\textsuperscript{\ref{footnote2}} & $9\times10^9~{\rm atoms/s}$   & \cite{Dieckmann1998}   \\
                   & $16~{\rm mK}$\textsuperscript{\ref{footnote1}}    & $16.9~\mu{\rm K}$ & $4.3\times10^9~{\rm atoms/s}$   & \cite{Wang2024}        \\
3D OM              & $70~\mu{\rm K}$  & -              & $1.3\times10^8~{\rm atoms/s}$ & \cite{Berthoud1999}    \\
\makecell[l]{2D MOT+\\Off-axis OM} & \makecell[l]{$886~\mu{\rm K}$\\$231~\mu{\rm K}$\footnote{The lowest temperature measured in our configuration.\label{footnote3}}} & $94~\mu{\rm K}$  & \makecell[l]{$4.9\times10^9~{\rm atoms/s}$\\$3.3\times10^8~{\rm atoms/s}$\textsuperscript{\ref{footnote3}}} & \makecell[l]{This\\work} \\
\end{tabular}
\end{ruledtabular}
\end{table}

The light shift due to the MOT beams is obtained from the relative change of the Ramsey central frequency as a function of the MOT laser power. The frequency shifts shown in Fig.~\ref{fig:dfonMOTP}(a) are summarized in Fig.~\ref{fig:dfonMOTP}(b), showing that the sensitivity of the center frequency to the MOT laser power is $-1.653\times10^{-3}~\mathrm{Hz/mW}$. The source-induced light shift at a typical MOT power ($300~\mathrm{mW}$) is $-0.51(4)~\mathrm{Hz}$. This is negligible compared with Raman- or CPT-induced shifts \cite{Pati2011,Wu2019,AbdelHafiz2020}. If the MOT power is stabilized at the $10^{-6}$ level \cite{Tricot2018}, the resulting fractional instability from this source is below $7.6\times10^{-17}$. The combination of continuous operation, low decoherence, and ultra-low light shift demonstrates that this compact beam source is highly suitable for precision clocks and interferometers.

Table~\ref{tab:table1} compares our source with representative single-cell continuous cold-atom beam sources. In the absence of both pushing beams and off-axis OM beams, a 2D MOT produces atomic beams with longitudinal temperatures on the order of several tens of kelvin. With the introduction of a pushing beam, conventional 3D MOT and 2D$^+$ MOT sources typically achieve longitudinal temperatures in the tens-of-millikelvin range. In contrast, by employing off-axis OM beams without any laser light collinear to the atomic beam, our setup generates a continuous cold atomic beam with longitudinal temperatures that are reduced by two to three orders of magnitude relative to conventional approaches.

In addition, the source-induced light shift of $-0.51(4)~{\rm Hz}$ is approximately 400 times smaller than the $-200~{\rm Hz}$ reported using previous light-shift suppression methods \cite{Huang2016}, while the measured decoherence rate $6.95(8)~{\rm s^{-1}}$ is lower than the $10\pm10~{\rm s^{-1}}$ reported in Ref. \cite{Kwolek2020}.

\section{Conclusions and Outlook}

We have demonstrated a compact, single-cell continuous cold-atom beam source that achieves simultaneous three-dimensional cooling within a $50~\mathrm{mm}$ region. The source delivers up to $4.9(5)\times10^9~\mathrm{atoms/s}$ at a mean velocity of $11.5~{\rm m/s}$, with transverse and longitudinal temperatures of $94(5)~\mu\mathrm{K}$ and $886(218)~\mu\mathrm{K}$, respectively. It can also operate at a longitudinal temperature as low as $231(65)~\mu\mathrm{K}$ at a mean velocity of $5.1~{\rm m/s}$, corresponding to a flux of $3.3(18)\times10^8~\mathrm{atoms/s}$. Raman-Ramsey measurements confirm ultra-low decoherence and light shift, yielding $90.85(30)\%$ fringe contrast at a $100~\mathrm{mm}$ separation and a light shift of $-0.51(4)~\mathrm{Hz}$. With these features, the source constitutes a practical building block for next-generation continuous cold-atom interferometers and clocks.

Recent advances in optical tweezer architectures increasingly rely on continuous cold-atom sources to enable scalable and continuously operated quantum simulation and computation \cite{Li2025,Chiu2025,Muniz2025}. The compact source demonstrated here provides a high-flux continuous atomic beam with strongly suppressed scattered light and reduced decoherence, offering a practical and compatible building block for such tweezer-based platforms.

In this work, near-resonant scattering light is suppressed by spatially blocking light from the cooling region. An alternative and complementary approach to mitigating scattering, demonstrated by Hu et al. \cite{Hu2025}, relies on spectrally shielding ground-state qubits by deliberately light-shifting the excited state. These distinct strategies offer different pathways for reducing decoherence in continuous cold-atom platforms.

The compact mirror structure used in this work is applicable to a broad range of cold-atom platforms. A contemporaneous and independent study by Nguyen et al. reports a similar geometry for generating a stationary cold atomic cloud\cite{Nguyen2025}, whereas our work focuses on a continuous cold atomic beam with intrinsic three-dimensional cooling and ultra-low light shift.

The mean atomic-beam velocity is tunable from $5$ to $20~\mathrm{m/s}$ by symmetrically shifting the OM laser frequencies $\delta_{\mathrm{OM}}$. The maximum achievable velocity and longitudinal temperature are limited by the OM interaction length, which determines the effective longitudinal cooling time. Higher velocities (up to $50~\mathrm{m/s}$) would further enhance the bandwidth and robustness of inertial sensors based on cold beams. This may be realized using two-color forward OM beams, analogous to the two-color pushing scheme employed in 2D$^+$ MOTs \cite{Park2012}.

Further improvements could reduce the temperature through polarization-gradient cooling, which could be suppressed by the residual magnetic field in the 2D MOT \cite{Riis1990,Rudolph2015,Kwolek2020}. The residual magnetic field in the cooling region is not an intrinsic limitation of the device. It arises from contributions such as the Earth's magnetic field, the cooling-cell metal structure, and the quadrupole magnetic field required for the operation of the 2D MOT in our configuration. In addition, tailoring the magnetic-field distribution to provide a polarization-gradient-cooling region inside the cooling cell may further improve the cooling performance. Other possible approaches include Raman sideband cooling \cite{Hamann1998} and dipole trapping \cite{Mashimo2019}.

\begin{acknowledgments}
The authors thank Dr. Jianwei Zhang (Department of Precision Instrument, Tsinghua University) for helpful discussions on light-shift measurements.
\end{acknowledgments}

\appendix

\section{Experimental Parameters}

This section summarizes the final experimental parameters used to achieve the results stated in the conclusion. Table~\ref{tab:table2} summarizes the laser-beam parameters-saturation parameters, polarizations, frequencies (referenced to the relevant atomic transitions), and beam sizes-used in the experiment. Unless otherwise stated, the reported atomic flux, transverse temperature, Raman-Ramsey contrast, and light shift are measured under regular operating conditions with a moving-OM frequency shift $\delta_{\rm OM} = \pm10~{\rm MHz}$, corresponding to an atomic velocity of approximately $11.5~{\rm m/s}$ and a longitudinal temperature of $886(12)~\mu\mathrm{K}$.

\begin{table*}[htbp]
\renewcommand{\arraystretch}{1.5}
\caption{\label{tab:table2}%
Laser-beam parameters under regular operating conditions of the cold-atom source.}
\begin{ruledtabular}
\begin{tabular}{lcclc}
\textrm{Laser}&
\begin{tabular}{c}
\textrm{Saturation parameter}\footnotemark[1]
\end{tabular}&
\textrm{Polarization}&
\makecell[c]{\textrm{Frequency}\footnotemark[2]}&
\textrm{Beam size}\\
\colrule
MOT       & 7.19  & $\sigma+$/$\sigma-$   & $(F=2\to F'=3)-4\Gamma$       & $50~{\rm mm}\times25~{\rm mm}$\\
Repumping & 0.72  & $\sigma+$/$\sigma-$   & $(F=1\to F'=2)$                 & $50~{\rm mm}\times25~{\rm mm}$\\
OM        & 3.53  & lin$\perp$lin         & $(F=2\to F'=3)-5\Gamma$ \footnotemark[3]  & $\phi~18~{\rm mm}$\\
State-Pre & 0.34  & lin$\parallel$lin     & $(F=2\to F'=2)\&(F=1\to F'=0)$  & $1~{\rm mm}\times20~{\rm mm}$\\
Probe     & 0.83  & $\sigma+$/$\sigma+$   & $(F=2\to F'=3)$                 & $1~{\rm mm}\times20~{\rm mm}$\\
\end{tabular}
\end{ruledtabular}
\footnotetext[1]{Laser intensity normalized to saturation intensity $I_{\rm sat}= 1.67 ~{\rm mW/cm^2}$}
\footnotetext[2]{Frequencies referenced to the corresponding atomic transitions hyperfine transitions; $\Gamma=2\pi\times6~\mathrm{MHz}$ denotes the natural linewidth.}
\footnotetext[3]{Average detuning between the forward and backward OM laser beams.}
\end{table*}

The lowest longitudinal temperature of $231(65)~\mu\mathrm{K}$ is obtained at a reduced mean atomic velocity of approximately $5~{\rm m/s}$, corresponding to $\delta_{\rm OM} = \pm3.5~{\rm MHz}$.

In addition, the quadrupole magnetic-field gradient associated with the 2D MOT is approximately $10~{\rm G/cm}$ along both the $x$ and $y$ directions.

\section{\label{app:simulation}Scattered light and ray-tracing simulation}
During continuous operation of the cold-atom beam source, near-resonant light is scattered as a result of spontaneous emission from the atoms and diffusive reflection of the cooling laser beams. Although no laser beam is collinear with the atomic beam, a fraction of this scattered light exits the cooling cell through the atomic output aperture, inducing light shifts and decoherence downstream.

In the ray-tracing simulation, atoms are modeled as an ensemble of point light sources with an emission power
\begin{equation}
P_{\rm sp}=E \times R_{\rm scatt}.
\end{equation}
where $E$ is the photon energy and $R_{\rm scatt}$ is the scattering rate given by Eq.~\ref{eq:scattering_force}, taking into account the atomic density distribution. The mirrors are modeled as secondary diffuse sources obeying Lambert' s law, accounting for the surface roughness and the optical power of the incident cooling beams. A Monte Carlo ray-tracing method is then used to propagate rays from these sources, with the atomic output aperture acting as a geometrical selector. The leaked optical power is obtained by counting the rays exiting through the aperture.

\nocite{*}

% \bibliography{apssamp}% Produces the bibliography via BibTeX.
\providecommand{\noopsort}[1]{}\providecommand{\singleletter}[1]{#1}%

\end{document}